\documentclass[12pt]{article}
  \usepackage{amsfonts}
\usepackage[nosort]{cite}
\usepackage{amsfonts}
\usepackage{amsmath}
\usepackage{amssymb}
\usepackage{amscd}
\usepackage{stmaryrd}
\usepackage{amssymb}
\usepackage{mathrsfs,eucal,graphicx}

\newcommand{\nn}{\nonumber}
\usepackage{array}
\usepackage{verbatim}
\begin{document}
\begin{flushright}

\end{flushright}

\vspace{1cm}

\begin{center}
{\Large\bf Maxwell symmetries and some applications} \\

\date{}
\vskip 3.0cm

{\bf Jos\'e A. de Azc\'arraga$^1$,
Kiyoshi Kamimura$^2$  and Jerzy Lukierski$^3$ } \\
\vskip 1.5cm
\small{ {$^1$ Department of Theoretical Physics,
University of Valencia and IFIC (CSIC-UVEG), 46100-Burjassot
(Valencia), Spain \\

$^2$ Department of Physics,
Toho University Funabashi, 274-8510,  Japan
\\

$^3$
Institute for Theoretical Physics,
University of Wroc{\l }aw, pl. Maxa Borna 9,
50--205 Wroc{\l }aw, Poland }
}
\vspace{2cm}

\end{center}

\begin{abstract}
The Maxwell algebra is the result of enlarging the Poincar\'{e} algebra
by six additional tensorial Abelian generators that make the fourmomenta
non-commutative. We present a local gauge theory based on the Maxwell algebra with vierbein,
spin connection and six additional geometric Abelian gauge fields. We apply this geometric
framework to the construction of  Maxwell gravity, which is described by the Einstein
action plus a generalized cosmological term. We mention a Friedman-Robertson-Walker
cosmological approximation to the Maxwell gravity  field equations, with
two scalar fields obtained from the additional gauge fields. Finally, we
outline further developments of the  Maxwell symmetries framework.
\end{abstract}


\vspace{1cm}

\newpage

\section{Introduction}

Maxwell symmetry was introduced around 40 years ago \cite{AKLM1,AKLM2},
but it is only recently that has attracted more attention. The $D=4$ Maxwell algebra, with
sixteen generators $(P_a, M_{ab}, Z_{ab})$, is obtained from Poincar\'{e} algebra if
we replace its commuting fourmomenta by noncommuting ones
\begin{equation}
[P_a, P_b]= \Lambda Z_{ab}\quad ,\quad [P_a, Z_{bc}]=0 \quad , \quad a,b=0,1,2,3\;,
\label{aklm1m1}
\end{equation}
where the six additional generators $Z_{ab}= -Z_{ba}$ are Abelian
and define a Lorentz-covariant tensor,
\begin{equation}
 [M_{ab}, Z_{cd}]=- (\eta_{c[a} Z_{b]d}- \eta_{d[a}Z_{b]c}) \; .
\label{aklm2m2}
\end{equation}
Further, we assume that the $Z_{ab}$ are dimensionless, which implies
that the parameter $\Lambda$ has mass dimensions $[\Lambda]=M^2$.
The Maxwell algebra $\mathcal{M}$ is the semidirect sum
${\mathcal{M}}= so(3, 1) \inplus {\mathcal{M}}^{\mathcal{I}}$,
where ${\mathcal{M}}^{\mathcal{I}}$ is the Maxwell ideal generated by
$P_a, Z_{ab}$ (eq.~(\ref{aklm1m1})), in which $\Lambda$ is the central
extension parameter. ${\mathcal{M}}$ can be obtained from
the $so(3, 2)$ algebra $(\mathcal{M}_{AB}; A, B=0,1,2,3,4)$ after the
rescaling  $\mathcal{M}_{ab}=\beta^2 Z_{ab}\,;\, a,b=0,1,2,3$,
$\mathcal{M}_{a4}=\beta \Lambda^{-\frac{1}{2}}P_a$ in the  contraction limit
$\beta\rightarrow \infty$ ($\beta$ is a dimensionless parameter).

The global Maxwell symmetries have been introduced in order to describe
Minkowski space with constant e.m. background\cite{AKLM1,AKLM2,AKLM3,AKLM4}
in models of relativistic particles interacting with a constant
e.m. field\footnote{For the non-relativistic case see
also \cite{CUP}, Sec.~8.3.}. In this paper, following \cite{AKLM5},
we present the construction of a local $D=4$ gauge theory
based on the Maxwell algebra (eqs.~(\ref{aklm1m1},\ref{aklm2m2}))
and apply it to generalize Einstein gravity. Such a theory will
accordingly contain six additional geo\-metric Abelian gauge fields,
playing the role of vectorial inflatons\footnote{For vector inflatons described
by $SU(2)$ gauge fields and non-abelian `gauge-flation' models
see e.g. refs.~\cite{AKLM6,AKLM7,AKLM8,AKLM9}.}
and which in Maxwell gravity contribute to a generalization of
the cosmological term. Further, we shall mention a one-dimensional FRW
cosmological approximation  describing the cosmic  scale factor $a(t)$ in
Maxwell gravity and comment briefly on other uses of
Maxwell symmetries.

\section{Gauging the Maxwell algebra to generalize Einstein gravity}

In order to introduce geometrically the Maxwell gauge vector fields
we consider the Maxwell algebra-valued one-form $h=h_\mu dx^\mu$, where
\begin{equation}
h_\mu= e^a_{\ \mu} P_a + \frac{1}{2}\omega^{ab}_{\ \ \mu} M_{ab} +
\frac{1}{2}A^{ab}_{\ \ \mu} \, Z_{ab} \; .
\label{aklm3m3}
\end{equation}
The Maxwell multiplet
$(e^a_{\ \mu}(x) , \omega^{ab}_{\ \ \mu}(x), A^{ab}_{\ \ \mu}(x))$
includes the vierbein, the spin connection and the new Abelian gauge fields
$A^{ab}_\mu$, which we interpret as geometrical inflaton vector fields.
Their associated curvatures are the components of the $\mathcal{M}$-valued
curvature two-form $R=R_{\mu\nu}dx^\mu\wedge dx^\nu$,
\begin{equation}
R_{\mu\nu}= T^a_{\ \mu\nu} P_a + \frac{1}{2}R^{ab}_{\ \ \mu\nu} M_{ab} +
\frac{1}{2}F^{ab}_{\ \ \mu\nu} \, Z_{ab} \; ,
\label{aklm4m4}
\end{equation}
which defines the two-forms corresponding to the
torsion $T^a$, the Lorentz curvature $R^{ab}$ and the field
strength $F^{ab}$ for $A^{ab}$, with dimensions
$[T]=M^{-1}\,,\,[R]=M^0=[F]$
(the one-forms $e^a=e^a{}_\mu dx\,,\,
\omega^{ab}=\omega^{ab}{}_\mu dx^\mu \,, \,
A^{ab}=A^{ab}{}_\mu dx^\mu$
have dimensions $[e^a]=M^{-1}\,,\, [\omega^{ab}]=M^0=[A^{ab}]$).
Explicitly, it follows from (\ref{aklm3m3})
and $R = dh+\frac{1}{2}[h,h]$ that the $T^a\,,\,R^{ab}\,,\,F^{ab}$
spacetime components are given by
\begin{eqnarray}
T^{a}_{\ \mu\nu} &=&\partial_{[\mu}\ e^{a}_{\ \nu]} +
\omega ^{a}_{\ b [\mu}\ e^{b}_{\ \nu]}\equiv D^{a}_{\ b [\mu}\ e^b_{\ \nu]}\,,
\label{aklm5am5a} \\
R^{ab}_{\ \ \mu\nu} &=&\partial_{[\mu}\ \omega^{ab}_{\ \ \nu]} +
\omega^a_{\ c [\mu}\ \omega^{cb}_{\ \ \nu]}=(D\omega^{ab})_{\mu\nu}=
- R^{ba}_{\ \ \mu\nu}\,,
\label{aklm5bm5b} \\
F^{ab}_{\mu\nu} &=&D^a_{c[\mu}\ A^{cb}_{\ \nu]} +
\Lambda \ e^a_{[\mu}\ e^b_{\nu]}\; ,
\label{aklm5cm5c}
\end{eqnarray}
where $D^a_{\ b\mu}=\delta^a_{\ b}\ \partial_\mu + \omega^a_{\ b\mu}$.
We observe that the torsion and the Lorentz curvature are the same as in
standard Einstein gravity, which is the particular choice of the
Einstein-Cartan gravity described by the Poincar\'{e} gauge theory;
the $\Lambda$-dependent additional term, which recalls the contribution
to $R^{ab}$ in (A)dS gravity, enters through the new gauge
curvature $F^{ab}$.

The following two geometric (metric independent) Lagrangian
densities were considered in detail in ref.~\cite{AKLM5},
namely

1) The Lagrangian density that leads to Einstein gravity
\begin{equation}
\label{aklm6m6}
\mathcal{L}_{2}=
-\frac{1}{2\kappa\Lambda }\varepsilon _{abcd}\,R^{ab}{\wedge }\,F^{cd} \, .
\end{equation}
Indeed, using
\begin{equation}
\label{aklma7a7}
\varepsilon _{abcd}\,R^{ab}{\wedge }\,(DA)^{cd}
=d\,(\varepsilon _{abcd}R^{ab}{\wedge }A^{cd})\;,
\end{equation}
it follows that, modulo the above boundary term,
the $\mathcal{L}_{2}$ in eq.~(\ref{aklm6m6})
gives the Einstein gravity Lagrangian $\mathcal{L}_E$,
\begin{equation}
\label{aklma8a8}
\mathcal{L}_{2} \simeq \mathcal{L}_{E}\equiv
-\frac{1}{2\kappa }\varepsilon _{abcd}\,R^{ab}{\wedge }e^{c}{\wedge }e^{d}\; .
\end{equation}

2) The generalized cosmological term.

 Let us recall that the Lagrangian density for the standard
geometric Einstein cosmological (EC) term, proportional
to the cosmological constant $\lambda$ $([\lambda]= M^2 $), is
\begin{equation}
\label{aklma9a9}
\mathcal{L}_{EC}=\frac{\lambda}{4\kappa }\varepsilon _{abcd}e^a{\wedge}e^b{\wedge }e^{c}{\wedge }e^{d}\; .
\end{equation}
The generalized cosmological term depends both on the
standard cosmological constant $\lambda$ and on the
parameter $\Lambda$ in eq.~(\ref{aklm1m1}). It is defined by
\begin{equation}
\label{aklma10a10}
\mathcal{L}_{C}=\frac{\lambda}{4\kappa\Lambda^2 }\,\varepsilon _{abcd}\,F^{ab}{\wedge }F^{cd}=
\mathcal{L}_{EC}+\Delta\mathcal{L}_{C}\; ,
\end{equation}
where the additional piece $\Delta\mathcal{L}_{C}$ with respect
to eq.~(\ref{aklma9a9}) is given by
\begin{equation}
\label{aklma11a11}
\Delta\mathcal{L}_{C}=
\frac{\lambda}{4\kappa \Lambda^2 }\,\varepsilon _{abcd}\,[(DA)^{ab}{\wedge }(DA)^{cd} +
2 \Lambda (DA)^{ab}{\wedge }e^{c}{\wedge }e^{d}]\; .
\end{equation}

Our basic Maxwell gravity action $\mathcal{L}_{M}$ is then given\cite{AKLM5}
by the Lagrangian density
\begin{equation}
\label{aklm12m12}
\mathcal{L}_{M}=
\mathcal{L}_{E}+\mathcal{L}_{C}=\mathcal{L}_{E}+\mathcal{L}_{EC}+\Delta\mathcal{L}_{C} \; ,
\end{equation}
which includes a new contribution, $\Delta\mathcal{L}_{C}$, to the standard
cosmological term (\ref{aklma9a9}).
By varying  with respect to the independent field variables
the following three equations of motion are obtained
\begin{equation}
\label{aklm13am13a}
\delta \omega ^{ab}:\qquad T^{[a}{\wedge }e^{b]}+
\frac{\lambda}{\Lambda^2}F^{[a}{}_{c}{\wedge } A^{c|b]}=0 \; ,
\end{equation}
\begin{equation}
\delta e^{a}:\qquad\varepsilon _{abcd}\,e^{b}\wedge \left( R^{cd}  -
\frac{\lambda}{\Lambda} F^{cd}\right) =0 \; ,   \label{aklm13bm13b}
\end{equation}
\begin{equation}
\delta A^{ab}:\qquad (De)^{[a}{\wedge }e^{b]}+\frac{1}{\Lambda} {R^{[a}}_{e}{\wedge }A^{e|b]}=0 \; .
\label{aklm13cm13c}
\end{equation}
It is useful to introduce a shifted Lorentz curvature by
\begin{equation}\label{aklm14m14}
J^{cd}= R^{cd}-\frac{\lambda}{\Lambda} F^{cd} \, .
\end{equation}
If we assume $J^{ab}=0$ the equations (\ref{aklm13am13a}) and (\ref{aklm13cm13c})
become identical, but  this assumption describes only special solutions of these
equations. In the general $J^{ab}\neq 0$ case,
the field equations (\ref{aklm13am13a}-\ref{aklm13cm13c})
can be rewritten in simpler form.  Using $J^{ab}$, one obtains
\begin{equation}
\delta \widetilde\omega ^{ab}:\qquad (\widetilde DJ)^{ab}\equiv d J^{ab} +
\widetilde\omega^{[a|c} J^{\ b]}_c = 0 \, ,\label{aklm15am15a}
\end{equation}
where $\widetilde\omega^{ab}=\omega^{ab}-\frac{\lambda}{\Lambda}A^{ab}$
is a shifted spin connection, and
\begin{equation}
\label{aklm15bm15b}
\delta e^{a}:\qquad\varepsilon _{abcd}\,e^{b}\wedge J^{cd} =0 \, ,
\end{equation}
\begin{equation}
\delta A^{de}:\qquad \varepsilon_{abc[d} J^{ab}\,A^c_{\ e]}=0\, .
\label{aklm15cm15c}
\end{equation}
Equation (\ref{aklm15bm15b}) is the generalization of
the Einstein equation.

   Passing from tangent to world spacetime indices
($J^{\mu\nu}$=$e_a^{\ \mu} e_b^{\ \nu} J^{ab}$
= $\frac{1}{2}J^{\mu\nu}_{\ \ \rho\sigma}dx^\rho\wedge dx^\sigma$,
$F^{\mu\nu}=e_a^{\ \mu} e_b^{\ \nu} F^{ab}$,  etc.),
eq.~(\ref{aklm15bm15b}) can be written as follows
\begin{equation}
{J^\mu}_\nu-\frac{1}{2}{\delta^\mu}_\nu\,J\equiv {R^\mu}_\nu-\frac{\lambda}{\Lambda}{F^\mu}_\nu-
\frac{1}{2}{\delta^\mu}_\nu(R-\frac{\lambda}{\Lambda}F)=0 \, ,
\label{Einst_1}\end{equation}
where $J^{\mu}_{\rho}\equiv{J^{\mu\nu}}_{\rho\nu}$, $J\equiv{J^{\mu}}_{\mu}$
and, similarly,
$R^{\mu}_{\rho}\equiv{R^{\mu\nu}}_{\rho\nu}\ ,\, R\equiv{R^{\mu}}_{\mu}\ , \,
F^{\mu}_{\rho}\equiv{F^{\mu\nu}}_{\rho\nu},\, F\equiv{F^{\mu}}_{\mu}$.
More explicitly, eq. (\ref{Einst_1}) can be written in a more
familiar form as
\begin{eqnarray}
&&{R^\mu}_\nu-\frac{1}2\,R\,{\delta^\mu}_\nu\,-3\,\lambda\,{\delta^\mu}_\nu = \qquad \qquad
\nn\\ &&\qquad =
\frac{\lambda}{\Lambda}\,\left({e_a}^\mu {e_b}^\sigma(D_{[\nu}A_{\sigma]})^{ab}-{\delta^\mu}_\nu
{e_a}^\rho {e_b}^\sigma(D_{\rho}A_{\sigma})^{ab}\right) \, .
\label{EinsteinJR57}
\end{eqnarray}
We see that the source added to the standard gravity equations
with cosmological constant $\lambda$  contains
linear contributions from the new gauge fields. The second term in the
rhs of (\ref{EinsteinJR57}) provides a field-dependent modification
of the cosmological constant at the lhs of the equation.

We wish to add that:

1)
Using equations
(\ref{aklm13am13a}) and (\ref{aklm13bm13b}), the spin
connection may be expressed as a function $\omega^{ab}(e, A)$
of the vierbein and the new gauge fields (for perturbative solutions
see \cite{AKLM5}, Appendix). In such a way we obtain the second
order formulation of Maxwell gravity, with independent fields
$e^{a}_{\, \mu}$ and $A^{ab}_{\ \mu}$.

2)
Using eqs.~(\ref{aklm13am13a}), (\ref{aklm13cm13c})
and the Bianchi identities for the $F^{ab}$ curvature, we obtain the free
field equation  for new gauge fields  $A^{ab}_{\  \mu}$,
\begin{equation}
\label{aklm20m20}
(D F)^{ab}=0 \; .
\end{equation}
Eq.~(\ref{aklm20m20}) can be modified by a source term if we
add to the Lagrangian (\ref{aklm12m12}) non-geometrical
contributions containing the fields $A^{ab}_{\  \mu}$ as e.g., a kinematical
term proportional to the density $F^{ab}\wedge \ast F_{ab}$,
similar to the free Lagrangian $-\frac{1}{2}F\wedge \ast F$ of
Maxwell electrodynamics.

3)
In order to estimate the effect of the new gauge fields on the dynamics of
the universe we have considered recently\cite{AKLM10bis}
a cosmological FRW approximation to the field equations of Maxwell gravity.
After introducing a function $a(t)$ describing the time dependence
of the cosmic scale factor
\begin{equation}
e^{a}_{\, \mu} (x) = (e^{a}_{\, i} (x),
e^{a}_{\, 0}(x)) \stackrel{FRW}{\longrightarrow}
 (\delta^{a}_{\, i} a(t), 0) \; ,
 \label{aklm25x}
\end{equation}
where $a=0,1,2,3$ and $i,j=1,2,3$, the six Abelian gauge
fields $A^{ab}_{\,   \mu}$ are approximated in terms of the
one-dimensional inflaton fields $\phi_1\,,\,\phi_2$, as follows
\begin{equation}
\begin{array}{l}
A^{rs}_{\ \mu} (x) = (A^{rs}_{\ i} (x), A^{rs}_{\ 0}(x)) \stackrel{FRW}{\longrightarrow}
(\epsilon^{\ rs}_{i} \, \phi_1 (t), 0)
 \, ,
\\ \\
A^{0r}_{\ \mu} (x) = (A^{0r}_{\ i} (x), A^{0r}_{\ 0}(x)) \stackrel{FRW}{\longrightarrow}
(\delta^{\ a}_{i} \, \phi_2 (t), 0)\; .
\label{aklm26x}
\end{array}
\end{equation}

The usual way of introducing the vector inflaton fields is
based on Yang-Mills gauge fields\cite{AKLM8,AKLM9} with internal symmetry
indices. In our case these internal indices are replaced by tangent spacetime indices,
and the three-dimensional tensors appearing in formula
(\ref{aklm26x}), $\epsilon^{\, rs}_{i}$ and $\delta^{a}_{\, i}$ (for $a=1,2,3$),
are genuine three-dimensional $so(3)$ tensors.

 \section{Outlook}

To conclude, we make the following comments:

a) The action defining Maxwell gravity  was chosen to
obtain a generalization of the cosmological term. Nevertheless,
as for the standard Einstein-Hilbert Lagrangian (\ref{aklma8a8}),
the action that follows from (\ref{aklm12m12}) is only invariant under local
Lorentz transformations and spacetime diffeomorphisms, not
under the full local Maxwell algebra.

We would like to mention at this point that
other Maxwell generalizations of Einstein gravity,
invariant under the local Abelian gauge symmetries associated with the
$Z_{ab}$ generators, have been proposed recently. The locally
Maxwell-invariant gravity model in ref.~\cite{AKLM10} contains a rather
controversial torsion squared term, and the Maxwell-invariant
extensions proposed in ref.~\cite{AKLM11} differ from the
Einstein Lagrangian (\ref{aklma8a8}) only by a topological
term i.e., the Einstein field equations remain unaltered.
We also note that another modification  of Einstein gravity
(see {ref.~\cite{AKLM12}}), obtained by gauging a deformation
of the Maxwell algebra\footnote{For the classification of the Maxwell algebra
deformations see ref.~\cite{AKLM14bis}.}, $so(3, 1)\oplus so(3, 2)$,
has been proposed recently\footnote{Such a model
can be called AdS-Maxwell gravity.}.
The action of the deformed Maxwell gravity in \cite{AKLM12}
is invariant under local deformed Maxwell transformations,
but in the contraction limit that leads to the Maxwell Lie algebra
the Abelian local Maxwell symmetries are also broken, as in our case.

b) The Maxwell symmetries have been generalized to Maxwell
super\-sym\-met\-ries\cite{AKLM13}; in particular, the
$N$-extended Maxwell superalgebras were recently described in detail
in \cite{AKLM14}. When $N$=1 one obtains three different models of
lowest dimensional Maxwell superalgebras, containing a pair
of two-component Weyl charges.

c) It is known that the higher spin (HS) free fields can be described as a free
field theory on enlarged, tensorial spaces which contains the $D$-dimensional
`physical' spacetime as a submanifold\cite{AKLM15,AKLM16,AKLM17,AKLM18,BAPV}. It turns out
that for $D$=4 the free HS fields can be obtained  from the first quantization of spinorial
particle model on a ten-dimensional tensorial space \cite{AKLM16,AKLM17,AKLM18}.
However, the ten-dimensional group manifold generated by the Maxwell ideal
${\mathcal{M}}^{\mathcal{I}}$ (eq.~(\ref{aklm1m1})) of the  Maxwell Lie algebra
also defines a ten-dimensional extended $D=4$ spacetime that we
call Maxwell $D=4$ tensorial space. One can consider as well a
spinorial particle model on this new Maxwell tensorial space which,
after first quantization, should also provide an infinite-dimensional multiplet
of $D=4$ HS fields; such a model is under consideration\cite{AKLM19}.
\vskip 1.2cm

\section*{\small{Acknowledgements}}
One of the authors (J.L.) would like to thank the organizers of the III Galilei-Guang-Xi
Meeting (Beijing, October 2011) for their warm hospitality. This paper is supported by a
research grant from the Spanish Ministry of Science and Innovation (FIS2008-01980)
and by the Polish NCN grant 2011/B/S12/03354.


\bigskip


\begin{thebibliography}{99}

\bibitem{AKLM1}
H.~Bacry, P.~Combe, J.L.~Richard, Nuovo Cim. \textbf{A67}, 267 (1970).

\bibitem{AKLM2}
R.~Schrader, Fortschr. Phys. \textbf{20}, 701 (1972).

\bibitem{AKLM3}
 S. Bonanos, J. Gomis, J. Phys.{\bf A43}, 015201 (2010)
 [arXiv:0812.4140v3 [hep-th]].

\bibitem{AKLM4}
S. Bonanos, J. Gomis, K. Kamimura, J. Lukierski,
J. Math. Phys. {\bf 51}, 102301 (2010)
[arXiv:1005.3714v2 [hep-th]].

\bibitem{CUP}
J.A.~de~Azc\'arraga and J.~M.~Izquierdo,
{\it Lie groups, Lie algebras, cohomolgy and some applications in Physics},
Camb. Univ. Press, 1985

\bibitem{AKLM5} J.A.~de~Azc\'arraga, K.~Kamimura, J.~Lukierski,
Phys.Rev. {\bf D83}, 12403 (2011)
[arXiv:1012.4402 [hep-th]].

\bibitem{AKLM6}
L.H.~Ford, Phys. Rev. \textbf{D40}, 967 (1989).

\bibitem{AKLM7}
A.~Golovnev, V.~Mukhanov, V.~Vanchurin,
J. Cosmol. Astropart. Phys. {\bf 08060}, 009 (2008)
[arXiv:0802.2068 [astro-ph]].

\bibitem{AKLM8}
A.~Maleknejad, M.M.~Sheikh-Jabbari, arXiv:1102.1513 [hep-ph].

\bibitem{AKLM9}
D.V.~Gal'tsov, E.A.~Davydov,
Proc. Steklov Inst. Math. {\bf 272}, 119 (2011)
[arXiv:1012.2861 [gr-qc]].

\bibitem{AKLM10bis}
A.~Borowiec, J.~Lukierski, M.~Woronowicz, in preparation.

\bibitem{AKLM10}
D.V.~Soroka, V.A.~Soroka,
Phys. Lett. {\bf B707}, 160 (2012)
[arXiv:1101.1591 [hep-th]].

\bibitem{AKLM11}
R.~Durka, J.~Kowalski-Glikman, M.~Szczachor,
 Mod. Phys. Lett. \textbf{A26}, 2689 (2011)
 [arXiv:1107.4728 [hep-th]].

\bibitem{AKLM12}
R.~Durka, J.~Kowalski-Glikman, arXiv:1110.6812 [hep-th].

\bibitem{AKLM14bis}
J.~Gomis, K.~Kamimura, J.~Lukierski,
JHEP {\bf 0908}, 039 (2009)
[arXiv:0906.4464 [hep-th]].

\bibitem{AKLM13}
S.~Bonanos, J.~Gomis, K.~Kamimura, J.~Lukierski,
Phys. Rev. Lett. \textbf{104}, 090401 (2010).

\bibitem{AKLM14}
K.~Kamimura, J.~Lukierski,
Phys. Lett. \textbf{B707}, 292 (2012)
[arXiv:1111.3598 [math-ph]].

\bibitem{AKLM15}
C.~Fronsdal,
{\it Massless particles, ortosymplectic symmetry and
another type of Kaluza-Klein theory}, Preprint UCLA/85/TEP/10, in
{\em Essays on Supersymmetry}, Reidel, 1986 (Mathematical Physics
Studies, v. 8), p.~163.

\bibitem{AKLM16}
I.~Bandos, J.~Lukierski, D.~Sorokin,
Phys. Rev. \textbf{D61}, 045002 (2000)
[hep-th/9907113].


\bibitem{AKLM17}
M. Vasiliev, Phys. Rev.
\textbf{D66}, 06606 (2002), hep-th/0106149;
Fortsch. Phys. {\bf 52}, 702 (2004)
[hep-th/0401177].

\bibitem{AKLM18}
M.~Plyushchay, D.~Sorokin, M.~Tsulaia,
JHEP {\bf 0304}, 013 (2003)
[hep-th/0301067].

\bibitem{BAPV}
I.~A.~Bandos, J.~A.~de Azc\'arraga, M.~Pic\'on and O.~Varela,
Phys. Rev. \textbf{D69}, 085007 (2004)
[hep-th/0307106].

\bibitem{AKLM19}
S.~Fedoruk, J.~Lukierski and D.~Sorokin, in preparation.

\end{thebibliography}
\end{document}